\documentclass[a4paper,12pt]{article}
\renewcommand{\baselinestretch}{1.2}
\usepackage{graphicx}
\usepackage{german}
\usepackage{ifthen}
\usepackage{amssymb}
\begin{document}
\newboolean{PDFAUSGABE}
\setboolean{PDFAUSGABE}{false}
%
%
\title{Quantenelektrodynamik: Nah- oder Fernwirkungstheorie?}
\author{Walter Smilga \\ Isardamm 135 d, D-82538 Geretsried 
\footnote{e-mail: wsmilga@compuserve.com}}
\maketitle 
\renewcommand{\baselinestretch}{1.2}

\begin{abstract}
\begin{sloppypar}
Der Zweiteilchen-Zustandsraum der St"orungsentwicklung zur 
elastischen Elektron-Elektron-Streuung wird betrachtet.
In der Feynmanschen Formulierung der Quantenelektrodynamik als 
Fernwirkungstheorie l"asst sich eine Beziehung zwischen 
Kopplungskonstante und dem Verh"altnis der Zustandsdichten des 
physikalischen Zweiteilchen-Zustandsraums einerseits und einer   
Zweiteilchen-Pro\-dukt\-darstellung andererseits herstellen. 
Diese Beziehung ist erf"ullt, wenn der physikalische Zustandsraum 
durch eine irreduzible Zweiteilchen-Darstellung der Poincar\'e-Gruppe
gebildet wird.
Die Konsequenzen dieser Beobachtung werden diskutiert.
\end{sloppypar}
\end{abstract}

\renewcommand{\baselinestretch}{1.0}

\hyphenation{ma-the-ma-ti-schen}
\hyphenation{eu-klid-schen}
\hyphenation{Quan-ten-elek-tro-dyma-mik}
\hyphenation{in-zwi-schen}
\hyphenation{phy-si-ka-li-scher}

\begin{quote}"`It has been a mystery ever since it was discovered more than fifty years ago, and all good theoretical physicists put this number up on their wall and worry about it."'
Richard P. Feynman \end{quote}

\section{Einleitung}

Richard Feynman formuliert in seinen grundlegenden Arbeiten zur Quanten\-elektrodynamik von 1949/50 diese als Fernwirkungstheorie mit retardierter Wechselwirkung. 
In der Einleitung zu der Arbeit \cite{rpf}, in der er die Feynman-Graphen einf"uhrt, bezeichnet er Nah- und Fernwirkungsformulierung als komplement"ar und gleichwertig, um sich dann aus didaktischen Gr"unden f"ur die Fernwirkungsformulierung zu entscheiden.
Dies scheint weitgehend in Vergessenheit geraten zu sein, denn heute interpretieren wir Feynman-Graphen, mit denen Feynman die Fernwirkungsformulierung veranschaulichte, ganz unbek"ummert im Sinne einer Nahwirkungstheorie.
Wir deuten interne Linien als Teilchen, und ignorieren dabei, dass Feynman zwar von "`virtuellen Teilchen"' spricht, damit aber "Ubergangsamplituden und nicht etwa Teilchenzust"ande meint.
Vollends zur Verwirrung tragen dann Darstellungen bei, die versuchen, eine reale, wenn auch kurzzeitige Existenz virtueller Teilchen "uber die Heisenbergsche Unbestimmtheitsrelation zu begr"unden.

Diese Begriffsverwirrung verlangt nach einer Kl"arung.
Zudem deutet sie darauf hin, dass sich hier auch nach 60 Jahren Quantenelektrodynamik noch unbeackertes Terrain befindet.
Und Sie werden sehen, dass es sich lohnt, hier etwas tiefer zu graben.

\section{Beziehung zwischen Zustandsraum und \\ Feinstrukturkonstante}

\begin{sloppypar}

Betrachten wir folgenden Beitrag zur S-Matrix der elastischen Elektron-Elektron-Streuung (M{\o}ller-Streuung) in niedrigster N"ahe\-rung. 
Er hat diese Struktur
\begin{equation}
\int d{\mathbf{p}}_1 d{\mathbf{p}}_2 \, d{\mathbf{k}}_1 d{\mathbf{k}}_2 \dots
\; {\bar{b}({\mathbf{p}}_1 + {\mathbf{k}}_1) \, \gamma^\mu \, b({\mathbf{p}}_1)} \,         
\; {\bar{b}({\mathbf{p}}_2 - {\mathbf{k}}_2) \, \gamma^\nu \, b({\mathbf{p}}_2)} \; 
\; a_\mu ({\mathbf{k}}_1)\,a^\dagger_\nu({\mathbf{k}}_2) \, .          \label{1-1}
\end{equation}
Wenn man die Photon-Operatoren kontrahiert, wird daraus dieser Ausdruck
\begin{equation}
\int 
d{\mathbf{p_1}}d{\mathbf{p_2}}\,d{\mathbf{k}}\,\dots\,
b^\dagger({\mathbf{p_1+k}})\, \gamma^\mu \, b({\mathbf{p_1}}) \;
b^\dagger({\mathbf{p_2-k}})\, \gamma_\mu \, b({\mathbf{p_2}})  \;. 
\label{1-1a}
\end{equation}
Er beschreibt den Austausch eines Impulsquants zwischen Elektron 1 und 2, und zwar ohne, dass hier noch ein Photon-Operator sichtbar ist.
Diesen Vorgang k"onnen wir daher mit Feynman zu Recht als eine direkte Fern\-wir\-kung zwischen den beiden Elektronen deuten.

Hier f"allt aber noch etwas anderes auf, das uns weiterf"uhren wird.
Das Integral "uber die Elektron-Impulse mit dem infinitesimalen Vo\-lu\-menelement $d{\mathbf{p}}_1 d{\mathbf{p}}_2$ erstreckt sich "uber das volle Vo\-lu\-men des Zwei\-teilchen-Impulsraums $\mathbb{R}^3 \times \mathbb{R}^3$.
Eigentlich sollten in einem abgeschlossenen Zweiteilchensystem, und ein solches liegt vor, wenn die Streuung elastisch ist, nur Zweiteilchenzust"ande ein und derselben \begin{em}irreduziblen\end{em} Darstellung der Poincar\'e-Gruppe vorkommen. 
F"ur eine solche Darstellung gilt diese Beziehung 
\begin{equation}
(p^0_1 + p^0_2)^2 - ({\mathbf{p}}_1 + {\mathbf{p}}_2)^2 = m_{eff}^2 \; , \label{1-2}
\end{equation}
wobei $m_{eff}$ die effektive Masse des Zweiteilchensystems ist. 
Diese Gleichung definiert offenkundig einen nicht-euklidischen Parameterraum. 

Dazu passt aber nicht das euklidische Volumenelement $d{\mathbf{p}}_1 d{\mathbf{p}}_2$.
Man sollte meinen, dass hier eine Jacobi-Determinante 
$J({\mathbf{p}}_1,{\mathbf{p}}_2)$ fehlt, die den nicht-euklidischen Parameterraum auf den euklidischen $\mathbb{R}^3 \times \mathbb{R}^3$ abbildet. 
Demzufolge w"are das Volumenelement zu ersetzen durch
\begin{equation}
J({\mathbf{p}}_1,{\mathbf{p}}_2)\,d{\mathbf{p}}_1 d{\mathbf{p}}_2 \;. \label{1-3}
\end{equation}

Andererseits wissen wir aber, dass vor dem Volumenelement die Kopplungskonstante $e^2$ stehen muss, um die St"orungsrechnung in "Ubereinstimmung mit dem Experiment zu bringen.
Dann sollte aber diese Beziehung 
\begin{equation}
J({\mathbf{p}}_1,{\mathbf{p}}_2)\, = \, e^2 = \, \alpha \;\;\; (\hbar = c = 1) 
\end{equation} 
zwischen dem experimentellen Wert der Feinstrukturkonstante $\alpha$ und der Jacobi-Determinante bestehen.
Und das ist hochinteressant, denn, wenn das so ist, dann sollte uns diese Beziehung in die Lage versetzen, durch Messung der Feinstrukturkonstante auf die Struktur des physikalischen Zustandsraums zu schlie{\ss}en.
Ich will daher zur "Uberpr"ufung dieser Beziehung die Jacobi-Determinante bestimmen, die unserem nicht-euklidischen Parameterraum (\ref{1-2}) zugeordnet ist. 
\end{sloppypar}

\section{Berechnung der Jacobi-Determinante}

Dazu muss ich auf einige Elemente aus der Theorie der Homogenen R"aume und der Ma{\ss}-Theorie zur"uckgreifen. 
F"ur das Folgende k"onnen Sie einen Homogenen Raum als einen Parameterraum mit einer gewissen Symmetrieeigenschaft verstehen. Diese Symmetrie wird durch die transitive Wirkung einer Lie-Gruppe ausgedr"uckt.

Ich konstruiere zun"achst die der Problemstellung entsprechende Lie-Grup\-pe.
F"ur die Impulse des ersten Elektrons steht der volle $\mathbb{R}^3$ zu Verf"u\-gung, f"ur die des zweiten auf Grund der Bedingung (\ref{1-2}) dann nur noch der $\mathbb{R}^2$. 
Somit l"asst sich der Zustandsraum durch Anwendung der Gruppe $SO(3,1) \times SO(2,1)$ auf einen ausgew"ahlten Zweiteilchen-Zustand erzeugen. 
Diese Gruppe kann man als Untergruppe in die Lie-Gruppe $SO(5,2)$ einbetten. 
Der $SO(5,2)$ wiederum l"asst sich die Faktorgruppe $SO(5,2) / (SO(5) \times SO(2))$ zuordnen. 
Diese Faktorgruppe bildet einen homogenen Raum, auf den die $SO(5,2)$ transitiv wirkt.
Er kann als 5-dimensionaler komplexer Vektorraum realisiert werden.
Dieser l"asst sich durch eine M"obius-Transformation\footnote{Eine sch"one Animation zur M"obius-Transformation findet sich unter \\ http://www.ima.umn.edu/$\sim$arnold/moebius/} 
konform und unter Erhaltung der Homogenit"atseigenschaft auf die komplexe Einheitskugel $D^5$ in f"unf Dimensionen abbilden.
Die komplexe Einheitskugel stellt damit eine nat"urliche Parametrisierung eines Zustandsraums mit einer $SO(5,2)$-Symmetrie dar.

Ich nehme im folgenden an, dass ich mein Integral (\ref{1-1a}) auf der Einheitskugel durch deren Koordinaten parametrisiert habe und integriere zun"achst "uber die Beitr"age auf der \begin{em}Oberfl"ache\end{em} $Q^5$ der Einheitskugel: 
\begin{equation}
\int_{Q^5} dQ^5 \dots = \int_{Q^5} d^4\xi \,/\, V(Q^5)  \dots \; . \label{1-4} 
\end{equation}
Dabei bezieht sich $\xi$ auf ein komplexes kartesisches Koordinatensystem einer Einbettung von $Q^5$ in den komplexen $\mathbb{C}^5$.
$V(Q^5)$ ist der Fl"acheninhalt von $Q^5$.
Jedem infinitesimalen Fl"achenelement ist nun eine gewisse Anzahl von Zust"anden, besser gesagt, eine Zustandsdichte, oder, mathematisch abstrakt, ein Ma{\ss} zugeordnet.
Diese Zust"ande erhalte ich, indem ich etwa auf einen Zustand in der Mitte des Fl"achenelements
infinitesimale Operationen der Gruppe $SO(5)$ anwende, die nicht "uber das Fl"achenelement hinausf"uhren.
Die "`Anzahl"' der Zust"ande entspricht somit der "`Anzahl"' dieser Gruppenelemente und diese definieren das Ma{\ss}, das dem Integral unter der Annahme einer $SO(5)$-Symmetrie zugrunde liegt.

Nun liegt aber tats"achlich keine volle $SO(5)$-Symmetrie vor, sondern nur eine solche bez"uglich der Untergruppe $SO(3) \times SO(2)$. 
Zur vollen $SO(5)$ fehlt die Drehung einer Impulskomponente des ersten Elektrons aus dem $\mathbb{R}^3$ in den $\mathbb{R}^2$ des zweiten Elektrons.
Eine solche Drehung tr"agt daher auch nicht zur "`Anzahl"' der Zust"ande im Fl"achenelement bei. 
Die Dichte der Zweiteilchenzust"ande ist demnach gegen"uber der vollen $SO(5)$-Symmetrie um das Volumen der Faktorgruppe $SO(5) / SO(4)$ reduziert.
Letztere ist darstellbar als Oberfl"ache $S^4$ der (reellen) Einheits\-kugel in f"unf Dimensionen. 
Das Ma{\ss} des Volumenelements in (\ref{1-4}) ist daher durch das Inverse des Volumens $V(S^4)$ zu korrigieren. Der Satz von Radon-Nikodym aus der Ma{\ss}theorie stellt sicher, dass dies eine mathematisch korrekte Vorgehensweise ist. 
Man erh"alt dann dieses korrigierte Volumenelement:
\begin{equation}
d^4\xi \, / \, ( V(Q^5) \, V(S^4) ) . \label{1-5}
\end{equation}

Ich kann auf der Oberfl"ache $Q^5$ noch "uber einen Phasenfaktor $e^{i\phi}$ integrieren, da der Integrand nicht explizit von $\phi$ abh"angt. 
(Diese Phase beschreibt die Orientierung des Zweiteilchen-Viererimpulses in der 
$p^0_1$-$p^0_2$-Ebene. Der Integrand h"angt aber nicht von $p^0_1$ und $p^0_2$ ab, da sich diese "uber die Massenbeziehung eliminieren lassen.)
Dies liefert einen zus"atzlichen Faktor von $2\pi$. 
Das Volumenelement erh"alt damit diese Form:
\begin{equation}
2 \pi \, d^4\xi \, / \, ( V(Q^5) \, V(S^4) ) \; , \label{1-51}
\end{equation}
wobei $\xi$ jetzt als reell zu betrachten ist.

Als n"achstes m"ochte ich dieses vierdimensionale Volumenelement zu einem \begin{em}f"unfdimensionalen kartesischen\end{em} Volumenelement erweitern. 
Ich greife dazu einen Punkt auf der Oberfl"ache $Q^5$ heraus und ordne zun"achst die vier kartesischen Koordinaten $\xi$ des Tangentialraums den vier Koordinaten des Impulsraums $\mathbb{R}^5$ zu. 
Die f"unfte Dimension erhalte ich dann, indem ich das Oberfl"achen\-element in radialer Richtung zu einem f"unf\-dimensionalen Volumenelement erweitere.
Das Volumenelement mu{\ss} isotrop in Bezug auf die f"unf orthogonalen Richtungen des $\mathbb{R}^5$ sein, denn ich will es ja im Integral (\ref{1-1a}) verwenden. 

Betrachten wir einmal das folgende Integral "uber das \begin{em}Volumen\end{em} der Einheitskugel $D^5$, ausgedr"uckt in sph"arischen Koordinaten: 
\begin{equation}
V(D^5) = 5 \int^1_0 dr \int_{Q^5} r^4 \, dQ^5 \; .    \label{1-6}
\end{equation}
Man kann dieses Volumen auch darstellen als ein Produkt von vier Integralen "uber jeweils eine Dimension des Oberfl"achenelements und ein f"unftes, wie vorher, "uber den Radius von $D^5$: 
\begin{equation}
5 \int^1_0 dr 
\int^r_0 V(D^5)^{\frac{1}{4}} dr'
\int^r_0 V(D^5)^{\frac{1}{4}} dr''
\int^r_0 V(D^5)^{\frac{1}{4}} dr'''
\int^r_0 V(D^5)^{\frac{1}{4}} dr''''  \; . \label{1-7}
\end{equation}
(Beweis durch Nachrechnen.)
Dieser Ausdruck stellt die "`Quadratur des Kreises"' dar, d.h. die Umwandlung eines sph"arischen Volumens in einen f"unfdimensionalen Quader gleichen Volumens 
\begin{equation}
1 \times V(D^5)^{\frac{1}{4}} \times V(D^5)^{\frac{1}{4}} \times V(D^5)^{\frac{1}{4}} \times V(D^5)^{\frac{1}{4}}\;. 
\end{equation}
Isotropie erfordert aber gleiche Kantenl"angen.
Um den Quader zu einen W"urfel umzuformen, mu{\ss} ich daher die Kante in radialer Richtung (und nur diese) durch eine Koordinatentransformation um den Faktor
\begin{equation}
V(D^5)^{\frac{1}{4}}                       \label{1-8}
\end{equation}
dehnen. 
Dieser Faktor ist die Jacobi-Determinante dieser Transformation.
Wenn ich diesen W"urfel dann in infinitesimale W"urfel zerlege, erhalte ich isotrope f"unfdimensionale Volumenelemente.
Mittels der Umkehrabbildung der M"obius-Trans\-for\-ma\-tion erhalte ich dann isotrope, kartesische Volumenelemente im $\mathbb{R}^5$.

Wenn man noch ber"ucksichtigt, dass "uber $2 \times 2$ Spin-Zust"ande zu summieren ist, 
erh"alt man einen zus"atzlichen Faktor 4, den ich in das Volumenelement einbeziehe.

Wenn ich alle Faktoren zusammenf"uge, erhalte ich diesen Volumenfaktor
\begin{equation}
8 \pi \,V(D^5)^{\frac{1}{4}} \, / \, (V(S^4) \, V(Q^5))   \label{1-9}
\end{equation}
als Jacobi-Determinante des infinitesimalen Volumenelement (\ref{1-3}). 
Nach Einsetzen der einzelnen Volumina \cite{lkh} erh"alt man diesen Wert
\begin{equation}
1 / 137,036 082 45 \;,
\end{equation}
der bis auf den Faktor $0,9999995$ mit der derzeit genauesten Messung von $e^2$, beziehungsweise $\alpha$, "ubereinstimmt \cite{hfg}.

Der Ausdruck (\ref{1-9}) wurde bereits vor 40 Jahren vom Schweizer Mathematiker Armand Wyler \cite{aw} gefunden. 
Allerdings konnte Wyler keine "uberzeugende Herleitung dieser Formel angeben. 
Insbesondere erkannte Wyler nicht den Zusammenhang mit den Zweiteilchen-Darstellungen der Poincar\'e-Gruppe. 
Der Bezug zur Physik blieb daher unklar.
Seine Formel wurde darum als reine Zahlenspielerei abgetan \cite{br}, zu Unrecht, wie sich jetzt zeigt.

\section{Schlu{\ss}folgerung}

\begin{sloppypar}
Was sagt uns dieses Ergebnis?

Es liefert zun"achst eine "uberraschend einfache Erkl"arung f"ur den numerischen Wert der Feinstrukturkonstante als derjenigen Jacobi-Determinante, die der Einbettung des physikalischen Zweiteilchen-Zustandsraums in den Zweiteilchen-Produktraum zugeordnet ist.

Gleichzeitig liefert dieses Ergebnis einen klaren experimentellen Hinweis darauf, dass in der nieder\-energetischen, elastischen Elektron-Streuung die beiden Elektronen in "au{\ss}erst guter N"aherung erstens ein abgeschlossenes System bilden und zweitens tats"achlich durch eine irreduzible Zweiteilchen-Darstellung beschrieben werden.
Ich habe diese Schlu{\ss}folgerung hier zwar nur anhand der niedrigsten N"aherung der Elektron-Streuung erl"autert; die h"oheren Terme der St"orungsreihe entstehen aber durch Iteration der niedrigsten N"aherung und tragen daher den gleichen Wert der Fein\-strukturkonstante.
Daher sind die gleichen Folgerungen auch f"ur alle h"oheren Ordnungen der St"orungsreihe zu ziehen. 

Damit ist der Zustandsraum der st"orungstheoretischen Quantenelektrodynamik in allen Ordnungen grunds"atzlich bekannt.
Er besteht demnach aus Zust"anden, die eindeutig als Zust"ande der beteiligten Teilchen zu identifizieren sind.
Dar"uberhinaus ist darin nichts enthalten, was mit einem weiteren eigenst"andigen Feld in Verbindung gebracht werden k"onnte.  
Damit ist offensichtlich, dass die Quantenelektrodynamik als relativistische Vielteilchen-Quantenmechanik mit retardierter Wechselwirkung im Sinne einer Fernwirkung  verstanden werden mu{\ss}. 
Zu kl"aren w"are noch, was die Ursache dieser Wechselwirkung ist, wenn es nicht mehr das elektromagnetische Feld ist.
Hierzu verweise ich auf den Anhang meines Vortrags, den Sie unter
http://arxiv.org/pdf/0912.5486 im Internet finden.

\end{sloppypar}

\section{Anhang}

Das Standardmodell betrachtet Eichfelder als Ursache der elektromagnetischen, schwachen und starken Wechselwirkung.
Es hat daher zun"achst den Anschein, als w"urde die im Vorstehenden vorgenommene "`Eliminierung"' des Photonfeldes im krassen Gegensatz zum Paradigma des Standardmodells stehen.
Ich m"ochte jetzt zeigen, dass dieses Paradigma tats"achlich eng mit der Irreduzibilit"at der Zweiteilchenzust"ande verkn"upft ist.

Fragen wir zun"achst nach der generellen Struktur von Zweiteilchen\-zu\-st"an\-den einer irreduziblen Zwei\-teil\-chen\-darstellung der Poincar\'e-Gruppe.
Ihr Zustandsraum ist gekennzeichnet durch die Eigenwerte des Operators $P = P^\mu P_\mu$, wobei $P_\mu$ der Operator des Viererimpulses ist, und des Operators $W = -W^\mu W_\mu$ mit 
$W_\sigma = \frac{1}{2} \epsilon_{\sigma \mu \nu \lambda} M^{\mu \nu}P^\lambda $, wobei 
$M^{\mu \nu}$ die "`Drehungen"' in der $x_\mu$-$x_\nu$-Ebene erzeugt.
Ein einzelner Zweiteilchenzustand l"asst sich dann durch seinen (r"aumlichen) Impuls und die Komponente des Drehimpulses in Richtung seines Impulses kennzeichnen. 
Damit ein Zweiteilchenzustand Eigenzustand des Drehimpulses ist, muss er neben einem reinen Produktzustand mit vorgegebenen Einzelimpulsen ${\mathbf{p_1}}$ und ${\mathbf{p_2}}$ auch alle hieraus durch Drehung um die genannte Achse hervorgehenden Produktzust"ande enthalten.
Die allgemeine Form von Zweiteilchen-Impulseigenzust"anden ist demnach ein "`verschr"ankter"' Zustand, gebildet aus Produktzust"anden mit gleichem Gesamtimpuls.

Die Formierung solcher verschr"ankter Zust"ande wird durch den standardm"a{\ss}igen Fockraum-Formalismus zun"achst nicht unterst"utzt.
Unsere Einsicht, dass Zweiteilchensysteme innerhalb der Quantenelektrodynamik durch eine
irreduzible Darstellung beschrieben werden, verlangt daher nach einer entsprechenden Anpassung des Formalismus. 
Man kann versuchen, diese Anpassung als "`St"orung"' der "`freien"' Dirac-Gleichung zu implementieren, die dann durch eine St"orungsrechnung zu behandeln w"are.
Dann m"ussen wir uns einen St"orungsterm zurechtlegen, der innerhalb einer St"orungsrechnung genau das folgende bewirkt: Ein einlaufender reiner Zweiteilchen-Produkt\-zu\-stand ist umzuwandeln in einen verschr"ankten Zweiteilchenzustand mit gleichem Gesamtimpuls; 
und dies muss genau dann geschehen, wenn zwei einzelne Teilchenzust"ande innerhalb der St"orungsrechnung zu einem Zweiteilchenzustand zusammengefasst werden.

Wie k"onnte ein solcher St"orungsterm aussehen?
Wenn Sie die Quantenelektrodynamik ansehen, finden Sie die Antwort.
Dort ist n"amlich ein derartiger Mechanismus bereits implementiert.
Die Zweiteilchen-S-Matrix (\ref{1-1a}) beschreibt, bei Anwendung auf einen einlaufenden reinen Produkt\-zu\-stand, dessen Umwandlung in einen verschr"ankten Zweiteilchenzustand, und bei weiterer Anwendung auf einen auslaufenden Produktzustand, die R"uckwandlung in einen reinen Produktzustand.
Dieser Mechanismus wird genau dann aktiviert, wenn in der S-Matrix (\ref{1-1}) die beiden Photon-Operatoren kontrahiert werden, was der Vereinigung beider Teilchen unter einem gemeinsamen Zweiteilchenzustand entspricht.
Der St"orungsterm, den ich ben"otige, um dieses Verhalten zu erreichen, ist das altbekannte quantisierte Vektorpotential, dessen Einf"ugen in die Dirac-Gleichung im Standardmodell durch das Paradigma der Eichinvarianz begr"undet wird.
Jetzt haben wir daf"ur eine ganz andere, aber mathematisch v"ollig transparente Begr"undung gefunden.
Und diese erkl"art, weshalb die Annahme einer lokalen Eichinvarianz in der Vergangenheit so erfolgreich war.
Dadurch wird das Paradigma der Eichinvarianz als heuristisches Prinzip best"atigt.

Was bleibt dann aber von dem Ph"anomen der elektromagnetischen Wechselwirkung?
Eigentlich nur die Einsicht, dass die elektromagnetische Wechselwirkung die sichtbare "Au{\ss}erung eines elementaren quantenmechanischen Prinzips darstellt, n"amlich, dass ein abgeschlossenes System durch eine irreduzible Darstellung der Poincar\'e-Gruppe beschrieben wird.

\renewcommand{\baselinestretch}{1.1}


\begin{thebibliography}{99}
\begin{sloppypar}

\bibitem{rpf} R. P. Feynman, 
Physical Review \bfseries76\mdseries, 769-789 (1949).

\bibitem{aw} A. Wyler,
C. R. Acad. Sc. Paris \bfseries271A\mdseries, 186-188 (1971).

\bibitem{br} B. Robertson, 
Phys. Rev. Lett. \bfseries27\mdseries, 1545 (1971).

\bibitem{lkh} L. K. Hua, \begin{em}Harmonic Analysis of Functions of
Several Complex Variables in the Classical Domains\end{em},
(American Mathematical Society, Providence, 1963).

\bibitem{hfg} D. Hanneke, S. Fogwell, G. Gabrielse,
Phys. Rev. Lett. \bfseries100\mdseries, 120801 (2008).

\end{sloppypar}
\end{thebibliography}
\end{document}